\documentclass[superscriptaddress,aps,preprintnumbers,amsmath,showpacs,amssymb,prl,nofootinbib,reprint]{revtex4-1}
\usepackage{bm, color} 
\usepackage{amssymb,amsfonts,slashed,amsthm,amsmath,graphicx, soul}
\usepackage{subfigure}
\usepackage{hyperref}
\usepackage{tikz,xcolor}

\begin{document}


\newcommand{\vev}[1]{ \left\langle {#1} \right\rangle }
\newcommand{\bra}[1]{ \langle {#1} | }
\newcommand{\ket}[1]{ | {#1} \rangle }
\newcommand{\eV}{ \ {\rm eV} }
\newcommand{\KeV}{ \ {\rm keV} }
\newcommand{\MeV}{\  {\rm MeV} }
\newcommand{\GeV}{\  {\rm GeV} }
\newcommand{\TeV}{\  {\rm TeV} }
\newcommand{\1}{\mbox{1}\hspace{-0.25em}\mbox{l}}
\newcommand{\Red}[1]{{\color{red} {#1}}}

\newcommand{\lmk}{\left(}  
\newcommand{\rmk}{\right)}
\newcommand{\lkk}{\left[}  
\newcommand{\rkk}{\right]}
\newcommand{\lhk}{\left \{ }  
\newcommand{\rhk}{\right \} }
\newcommand{\del}{\partial}  
\newcommand{\la}{\left\langle} 
\newcommand{\ra}{\right\rangle}
\newcommand{\half}{\frac{1}{2}}

\newcommand{\bea}{\begin{array}}
\newcommand{\eea}{\end{array}}
\newcommand{\beq}{\begin{eqnarray}}
\newcommand{\eeq}{\end{eqnarray}}
\newcommand{\eq}[1]{Eq.~(\ref{#1})}

\newcommand{\dd}{\mathrm{d}}
\newcommand{\Mpl}{M_{\rm Pl}}
\newcommand{\mg}{m_{3/2}}
\newcommand{\abs}[1]{\left\vert {#1} \right\vert}
\newcommand{\mphi}{m_{\phi}}
\newcommand{\Hz}{\ {\rm Hz}}
\newcommand{\for}{\quad \text{for }}
\newcommand{\Min}{\text{Min}}
\newcommand{\Max}{\text{Max}}
\newcommand{\Kahler}{K\"{a}hler }
\newcommand{\cphi}{\varphi}
\newcommand{\Tr}{\text{Tr}}
\newcommand{\diag}{{\rm diag}}

\newcommand{\SUf}{SU(3)_{\rm f}}
\newcommand{\Upq}{U(1)_{\rm PQ}}
\newcommand{\Zpq}{Z^{\rm PQ}_3}
\newcommand{\Cpq}{C_{\rm PQ}}
\newcommand{\ubar}{u^c}
\newcommand{\dbar}{d^c}
\newcommand{\ebar}{e^c}
\newcommand{\nubar}{\nu^c}
\newcommand{\Ndw}{N_{\rm DW}}
\newcommand{\Fpq}{F_{\rm PQ}}
\newcommand{\fpq}{v_{\rm PQ}}
\newcommand{\Br}{{\rm Br}}
\newcommand{\Lag}{\mathcal{L}}
\newcommand{\Lqcd}{\Lambda_{\rm QCD}}

\newcommand{\ji}{j_{\rm inf}} 
\newcommand{\jb}{j_{B-L}} 
\newcommand{\M}{M} 
\newcommand{\im}{{\rm Im} }
\newcommand{\re}{{\rm Re} }

\def\lrf#1#2{ \left(\frac{#1}{#2}\right)}
\def\lrfp#1#2#3{ \left(\frac{#1}{#2} \right)^{#3}}
\def\lrp#1#2{\left( #1 \right)^{#2}}
\def\REF#1{Ref.~\cite{#1}}
\def\SEC#1{Sec.~\ref{#1}}
\def\FIG#1{Fig.~\ref{#1}}
\def\EQ#1{Eq.~(\ref{#1})}
\def\EQS#1{Eqs.~(\ref{#1})}
\def\TEV#1{10^{#1}{\rm\,TeV}}
\def\GEV#1{10^{#1}{\rm\,GeV}}
\def\MEV#1{10^{#1}{\rm\,MeV}}
\def\KEV#1{10^{#1}{\rm\,keV}}
\def\blue#1{\textcolor{blue}{#1}}
\def\red#1{\textcolor{red}{#1}}

\newcommand{\eff}{\Delta N_{\rm eff}}
\newcommand{\neff}{\Delta N_{\rm eff}}
\newcommand{\cc}{\Lambda}
\newcommand{\Mpc}{\ {\rm Mpc}}
\newcommand{\Msolar}{M_\cdot}


\title{
Abelian-Higgs vortices in the oscillating axion background
}

\author{Naoya Kitajima}
\affiliation{Frontier Research Institute for Interdisciplinary Sciences, Tohoku University, \\
6-3 Azaaoba, Aramaki, Aoba-ku, Sendai 980-8578, Japan}
\affiliation{Department of Physics, Tohoku University, \\
6-3 Azaaoba, Aramaki, Aoba-ku, Sendai 980-8578, Japan}

\author{Shota Nakagawa}
\affiliation{Tsung-Dao Lee Institute, Shanghai Jiao Tong University, \\
No.~1 Lisuo Road, Pudong New Area, Shanghai 201210, China}
\affiliation{School of Physics and Astronomy, Shanghai Jiao Tong University, \\
800 Dongchuan Road, Shanghai 200240, China}

\begin{abstract}
We study the dynamics of Abelian-Higgs vortices in the background of a coherently oscillating axion field.
We show that the electric field is induced in the magnetic core of the vortex due to the axion-photon conversion.
Moreover, because the electromagnetic field is confined in the vortex and excluded from the superconducting bulk regions due to the Meissner effect, the vortex tube can be regarded as a cylindrical cavity, and our numerical analysis shows that the resonant cavity mode (TM010 mode) can be efficiently enhanced in this tube. We also focus on the interaction of two vortices in the oscillating axion background, resulting in attractive or repulsive forces, even in the case with the BPS limit. These new features open up a new possibility for the axion dark matter search using superconducting devices.
\end{abstract}

\maketitle
\flushbottom

{\bf Introduction.--}
Axion is one of the most plausible candidates for dark matter.
While an axion, known as the QCD axion, plays an essential role in the Peccei-Quinn mechanism \cite{Peccei:1977hh,Peccei:1977ur,Weinberg:1977ma,Wilczek:1977pj}, a large number of axions also appear in low-energy effective theory of string theory \cite{Svrcek:2006yi,Arvanitaki:2009fg,Cicoli:2012sz}. 
See \cite{Sikivie:2006ni,Kawasaki:2013ae,Marsh:2015xka,DiLuzio:2020wdo,Choi:2020rgn,OHare:2024nmr} for reviews.
The axion is typically very light and behaves as wave-like dark matter \cite{Ferreira:2020fam,Hui:2021tkt}, so that it can be regarded as a coherent oscillation in table-top experiments (see \cite{Irastorza:2018dyq} for a review).
In particular, since the axion generally interacts with the photon through electromagnetic anomaly, 
a system permeated by strong external magnetic field is a good avenue to probe the axion \cite{Sikivie:1983ip}.
In fact, the currently most sensitive experiments use resonant cavities,
in which the electric and magnetic fields are amplified via the axion-photon conversion when the resonant frequency corresponds to the axion mass \cite{Brubaker:2016ktl,ADMX:2018gho,Alesini:2019ajt,ADMX:2019uok,Alesini:2020vny,HAYSTAC:2020kwv,Grenet:2021vbb,Adair:2022rtw,TASEH:2022vvu,CAPP:2024dtx,HAYSTAC:2024jch}.
Due to its feeble interaction, such an elaborated approach is required for the axion detection.

In this Letter, we find a novel perspective of the interplay between the axion dark matter and photons in Abelian-Higgs model, which will lead to a peculiar signature of the existence of the axion.
As a remarkable feature, the Abelian-Higgs model has a static solution known as a vortex, characterized by an integer value corresponding to the nontrivial topological structure of the vacuum.
The energy of the vortex is localized at the core, where the U(1) gauge symmetry is locally restored, in other words, the magnetic field is trapped inside the vortex with the quantized magnetic flux.
Specifically, this can be realized in a type-II superconductor.
The magnetic field can penetrate the bulk region with the superconducting state through the vortices, while it is excluded by the Meissner effect outside them.
In the presence of the axion oscillation inside the vortex, oscillating electromagnetic fields are generated due to the axion-photon conversion.\footnote{Ref. \cite{Iwazaki:2020agl} focuses on a similar situation, in which the oscillating electric field is induced by the axion on the surface of the superconductor with the applied magnetic field.}
Furthermore, as shown below, our numerical simulation observes that the induced electric field can get amplified in the vortex if the oscillation frequency is tuned to some resonant frequency corresponding to the vortex core size, implying that the vortex can be regarded as a cylindrical resonant cavity.

We also focus on the interaction between two vortices in the oscillating axion background.
So far, the interaction between vortices/cosmic strings has been studied in \cite{Jacobs:1978ch,Taubes:1979tm,Ruback:1988ba,Moriarty:1988fx,Myers:1991yh,Samols:1991ne,Rebbi:1991wk,Speight:1996px,Eto:2022hyt,Fujikura:2023lil}, verifying that the physical property is determined by the ratio $\beta\equiv\lambda/(2e^2)$ with $\lambda$ and $e$, the scalar self-coupling constant and the gauge coupling constant respectively (see Eq.~(\ref{Lag}) and below).
In other words, the interaction is determined by the mass ratio between the scalar field and the gauge field, or the ratio between the correlation length (scalar core size) and the magnetic penetration length (vector core size).
Namely, $\beta < 1$ ($\beta > 1$), corresponding to the type-I (II) superconductor/cosmic string, results in the attractive (repulsive) interaction between two aligned vortices/cosmic strings.
In the special case with $\beta = 1$, known as the critical coupling or Bogomol’nyi-Prasad-Sommerfield (BPS) limit \cite{Bogomolny:1975de,Prasad:1975kr}, no force is exerted between two aligned vortices.

In this Letter, we show that the above feature of vortex interactions can be significantly modified in the oscillating axion background.
The effect of the axion on the vortex dynamics has been studied in the literature \cite{JACOBS1988288}, where the author focuses on the static but spatially inhomogeneous axion fields.
In contrast, we focus on the homogeneous but time-dependent axion fields, bearing in mind the wave-like nature of the axion dark matter.

\vspace{2mm}
{\bf The Model.--}
Let us begin with an Abelian-Higgs model or Ginzburg-Landau (GL) model with an axion field, $a$, interacting with gauge fields \cite{Kitajima:2021bjq,Nakagawa:2022knn}. 
The model contains a complex Higgs scalar field or GL order parameter, $\Phi$, gauge fields, $A_\mu$, and the relativistic form of the Lagrangian is given by
\begin{align}
\mathcal{L} &= (D_\mu \Phi)^\dagger D^\mu \Phi -V_\Phi(\Phi) - \frac{1}{4} F_{\mu\nu}F^{\mu\nu}\nonumber\\
&+ \frac{1}{2}\del_\mu a\del^\mu a -V_a(a) -\frac{g_{a\gamma}}{4} a F_{\rm \mu\nu}\tilde{F}^{\mu\nu}.
\label{Lag}
\end{align}
Here $D_\mu = \partial_\mu - ieA_\mu$ is the gauge covariant derivative, $F_{\mu\nu} = \partial_\mu A_\nu - \partial_\nu A_\mu$ is the field strength tensor, $\tilde{F}^{\mu\nu} = \epsilon^{\mu\nu\rho\sigma} F_{\rho\sigma}/2$ is its dual, 
and the axion couples to gauge fields with a dimensionful coupling constant $g_{a\gamma}=c_\gamma\alpha/(\pi f_a)$, where $f_a$ denotes the axion decay constant, $c_\gamma$ the anomaly coefficient for the U$(1)$ gauge symmetry, and $\alpha\equiv e^2/(4\pi)$.
We consider the Mexican-hat potential for the Higgs field,
\begin{align} \label{eq:potential}
    V_\Phi(\Phi) = \frac{\lambda}{4} (|\Phi|^2 - v^2)^2,
\end{align}
with $\lambda$ being the self-coupling constant and $v$ the vacuum expectation value.
The axion acquires a periodic potential $V_a$ via nonperturbative effects,
and in the case of our interest, we assume $a \ll f_a$, then, the potential can be well-approximated as the quadratic potential with the axion mass $m_a$.
As long as the axion is light enough, the axion behaves as a coherent oscillation or nonrelativistic matter, typically produced via the misalignment mechanism \cite{Preskill:1982cy,Abbott:1982af,Dine:1982ah} and also the decay of topological defects \cite{Davis:1986xc,Lyth:1991bb} (see also e.g. \cite{Hiramatsu:2010yn,Hiramatsu:2010yu,Hiramatsu:2012gg,Hiramatsu:2012sc,Kawasaki:2014sqa}).
Specifically, we assume that the axion is a spatially homogeneous coherent oscillation given by
\begin{align} \label{eq:axion_osc}
a = \mathcal{A} \cos(\omega_a t + \delta),
\end{align}
where $\mathcal{A}$ and $\omega_a( \approx m_a)$ are the amplitude and frequency of the axion oscillation respectively, and $\delta$ is the initial phase. 
This simplification is justified when we consider the physical system of superconducting devices whose typical size is much smaller than the de Broglie wave length of the axion dark matter.
Hereafter, we set $\delta=0$ without loss of generality, and $g_{a\gamma}$ is absorbed by the definition of $\mathcal{A}$.

The potential (\ref{eq:potential}) has a vacuum manifold of $S^2$ and admits a static solution known as the Abrikosov-Nielsen-Olesen (ANO) vortex \cite{Abrikosov:1957wnz,Nielsen:1973cs}.
In particular, when the U$(1)$ gauge symmetry is spontaneously broken, vortices (or cosmic strings) are copiously produced. Note that the mass of the Higgs and gauge fields are given by, respectively, $m_\Phi = \sqrt{\lambda}v$ and $m_A = \sqrt{2}ev$ in the broken symmetry phase, while the symmetry is restored inside the vortex and the gauge fields are massless there, allowing the existence of stable magnetic flux.
Since the dark matter axion is spatially ubiquitous, 
we will find that the axion coherent oscillation
in the vortex leads to some nontrivial phenomena through the axion-photon conversion.

From the Lagrangian (\ref{Lag}),
the equations of motion for the Higgs field and the gauge fields and the constraint equation in the Minkowski spacetime are given by
\begin{align}
& \ddot{\Phi} - D_i D_i \Phi +\frac{\del V_\Phi}{\del\Phi^*} = 0,
\label{Higgs}\\
& \dot{E}_i +\nabla^2A_i -\del_i\del_jA_j + 2e{\rm Im}[\Phi^* D_i\Phi] \nonumber\\ & -g_{a\gamma}(\dot{a} B_i- \epsilon_{ijk} (\del_ja)E_k) = 0,
\label{Ampere} \\
& \del_i E_i + 2e{\rm{Im}}[\Phi^*\dot{\Phi}] + g_{a\gamma}(\del_ia) B_i = 0,
\label{Gauss}
\end{align}
where we take the temporal gauge, $A_0=0$, $E_i\equiv F_{i0}$ and $B_i \equiv \epsilon_{ijk} \partial_j A_k$ are, respectively, the electric and the magnetic fields, and the overdot represents the time derivative.
In what follows, we substitute \EQ{eq:axion_osc} for the axion field and we drop the spatial gradient term, i.e. $\partial_i a = 0$.
Note that the dynamics is determined by only one parameter, $\beta = \lambda/(2e^2)$, by properly rescaling the spacetime coordinates and the field values.

\begin{figure}[tp]
\centering
\subfigure{\includegraphics[width = 8.5cm, clip]{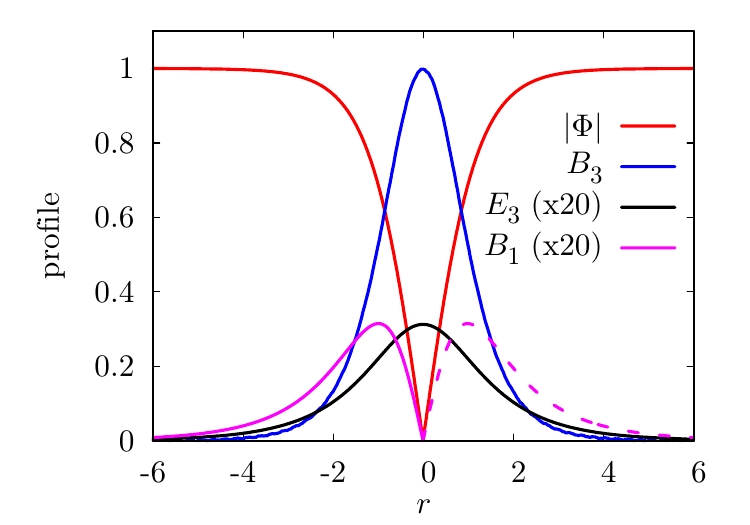}}
\label{subfig:profile}
\subfigure{\includegraphics [width = 8.5cm, clip]{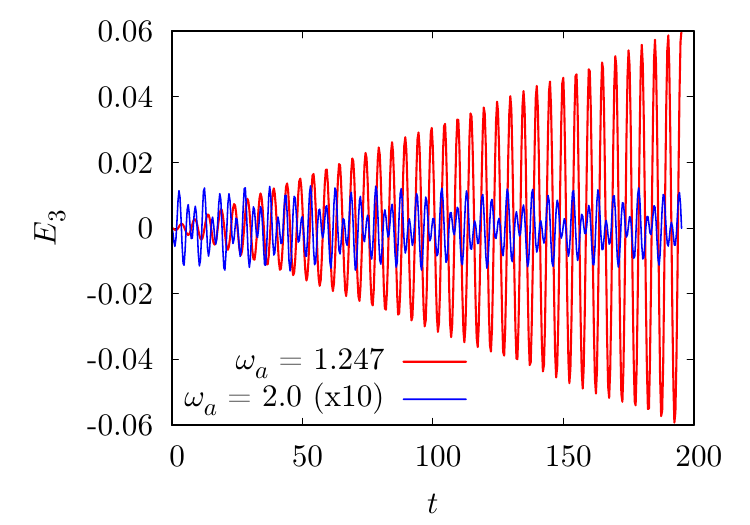}}
\label{subfig:evolve}
\caption{
Top: The profile of the vortex solution for $\beta=1$ along the $y$-axis at $t = 197$.
The red, blue, black, and magenta lines correspond respectively the radial component of the Higgs field, the magnetic field along the $z$-axis, the induced electric field along the $z$-axis, and the induced magnetic field along the $x$-axis. The amplitude and the frequency of the axion are $\mathcal{A} = 0.001$ and $\omega_a = 1.247$.
The dashed line represents the negative value, and $E_3$ and $B_1$ are multiplied by $20$ for the illustration purpose.
Bottom: Time evolution of the induced electric field for $\omega_a = 1.247$ (red) and  2.0 (blue, multiplied by a factor 10) with the axion amplitude $\mathcal{A} = 0.001$.
}
\label{fig:single_vortex}
\end{figure}

One can find a static ANO vortex solution for \EQ{Higgs} and \EQ{Ampere} by taking the following ansatz in the cylindrical coordinate system $(r,\theta,z)$,
\begin{align}
& \Phi = v f(r) e^{i n \theta}, \label{eq:vortex_Higgs}\\
& A_r = A_z = 0,~~ A_\theta = -\frac{n v \alpha(r)}{er},\label{eq:vortex_gauge}
\end{align}
with boundary conditions, $f(\infty)=\alpha(\infty) = 1$, and $f(0)= \alpha(0) = 0$.
Here, $n$ is an integer called the winding number.
One can obtain the spatial profiles by solving the following equations,
\begin{align}
&f'' + \frac{f'}{r}-\frac{n^2(\alpha-1)^2 f}{r^2} - \frac{\lambda f(f^2 - 1)}{2} =0,\label{vortex_sol2}\\
&\alpha'' - \frac{\alpha'}{r} -2e^2 f^2 (\alpha - 1) = 0,
\label{vortex_sol1}
\end{align}
where the prime denotes the derivative with respect to $r$.
Note that we ignore the contribution from the axion-photon interaction in the above equations.
The profiles can be calculated numerically as shown in the upper panel of Fig. \ref{fig:single_vortex}.

\vspace{2mm}
{\bf Oscillating axions in the vortex core.--}
We study the influence of the coherently oscillating axions in the vortices. 
Let us approximately identify the vortex 
as a cylindrical cavity whose interior is assumed to be empty vacuum.
A more realistic consideration will be required to demonstrate the actual behavior of the axion with experimental setups, 
but the present paper points out a novel aspect of the vortex physics, 
so such a simplified assumption seems suitable as a first step.
We will discuss future prospects on this point in the Discussion section.

\EQ{Ampere} tells us that the oscillating axion in a region permeated by static magnetic fields induces oscillating electric fields with the same frequency as that of the axion.
In addition, such oscillating electric fields can be resonantly enhanced in a cavity if the frequency is tuned to the resonant frequency.
Since the static magnetic field penetrates the vortices, it is expected that we observe similar resonance behavior, which will be shown shortly.

\begin{figure}[tp]
\centering
\includegraphics [width = 8.5cm, clip]{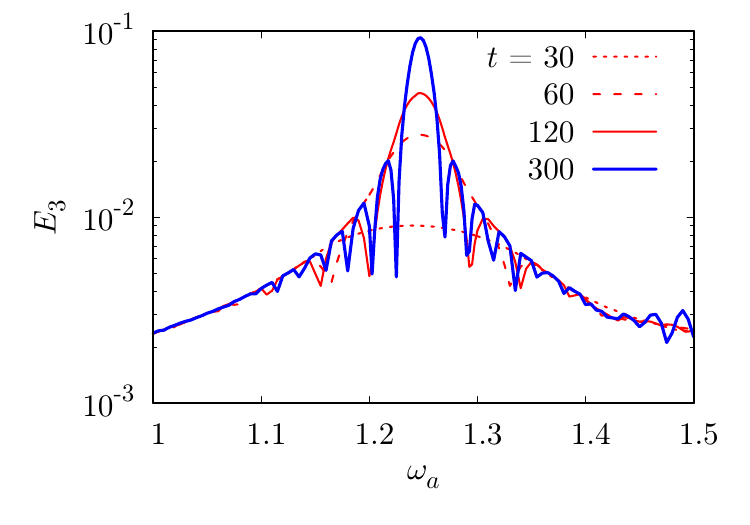}
\caption{
Amplitude of the induced electric field as a function of the axion frequency
at $t = 30, 90, 150, 300$ from bottom (dotted line) to top (blue solid line).
We have taken the axion amplitude $\mathcal{A} = 0.001$. 
}
\label{fig:resonance}
\end{figure}

In the cylindrical cavity, transverse magnetic modes, TM$_{0l0}$, can be coupled to the axion, with $l$ the number of nodes in the radial direction. 
The resonance frequency is given by
\beq
\omega_{{\rm TM}_{0l0}} =  \frac{\xi_l}{R},
\eeq
where $\xi_l$ is defined as the $l$-th root of the Bessel function of order 0, $J_0(x)=0$, 
and $R$ is the radius of the cylinder.
For the most dominant mode, TM$_{010}$ with $\xi_1\simeq 2.405$,
the induced electric field has a resonance peak at the axion frequency $\omega_a~(= m_a)$, 
\beq
\omega_a = \frac{\xi_1}{R_{\rm eff}},
\eeq
where $R_{\rm eff}$ is the effective cavity radius, which is the order of the magnetic core size $m_A^{-1}$.
From the numerical calculation, we found the resonant frequency, $\omega_a = 1.247v$ (see \FIG{fig:resonance}), and the effective cavity radius is obtained as $R_{\rm eff} \simeq 2.7m_A^{-1}$.

\vspace{2mm}
{\bf Numerical results.--}
We have performed numerical lattice simulations to solve the system of equations, \EQS{Higgs},(\ref{Ampere}), with the oscillating axion background \EQ{eq:axion_osc} based on the lattice gauge formulation \cite{Moriarty:1988fx}.
Implicitly assuming an experimental setup with a thin superconducting film, we consider the dynamics of aligned vortex lines. 
In this case, the spacetime can be practically regarded as $2+1$ dimensional. 
In other words, there is a shift symmetry along the direction of the vortex line which we identify the third spatial direction ($z$-direction).
In our simulations, we set the grid number $N^2 = 4096^2$, the boxsize $L = 64/v$ and the stepsize for time evolution $0.2 \delta x$ with $\delta x = L/N$ the lattice spacing. As a consistency check, we also performed simulations with different grid points and boxsizes, but we obtained the same results.
The initial values are given by \EQS{eq:vortex_Higgs},(\ref{eq:vortex_gauge}) with the numerical solutions of \EQS{vortex_sol2},(\ref{vortex_sol1}).
In what follows, all dimensionful physical quantities are normalized by $v$.

First, let us show the resonant amplification in the static single vortex. 
The upper panel of \FIG{fig:single_vortex} shows the profile of the single vortex solution ($|\Phi|$, $B_3$) and the induced electromagnetic fields ($E_3$, $B_1$).
One can see from the figure that the electric and the magnetic fields are induced with the direction along the core magnetic field and the circumference direction. 
The lower panel of \FIG{fig:single_vortex} shows that the amplitude of the induced electric field is linearly enhanced for $\omega_a = 1.247$, showing the feature of forced oscillation for resonant cavity modes.
\FIG{fig:resonance} shows the amplification of the electric field as a function of the axion frequency with the amplitude $\mathcal{A} = 0.001$.
This figure has a sharp peak and clearly shows that there is a resonant frequency as in the case of the resonant cavity.

Next let us consider the dynamics of two vortices to clarify the influence of the axion-induced electromagnetic fields. 
Both vortices are initially at rest and we use the following ansatz for the initial field values \cite{Vilenkin:2000jqa},
\begin{align}
    \Phi(\bm{x}) &= \Phi^{(1)}(|\bm{x}-\bm{x}_1|) \Phi^{(2)}(|\bm{x}-\bm{x}_2|),\\
    A_i(\bm{x}) &= A_{i}^{(1)}(|\bm{x}-\bm{x}_1|) + A_{i}^{(2)}(|\bm{x}-\bm{x}_2|),
\end{align}
where $\Phi^{(s)}$ and $A_i^{(s)}$ ($s=1,2$) are the single vortex solutions for the Higgs and the gauge fields, centered at $\bm{x} = \bm{x}_{s}$.
Our simulations show that the vortices are unmoved in the BPS limit without the axion.
On the other hand, when we turn on the axion oscillation, the vortices start to move in a finite time.

\begin{figure}[tp]
\centering
\includegraphics [width = 8.5cm, clip]{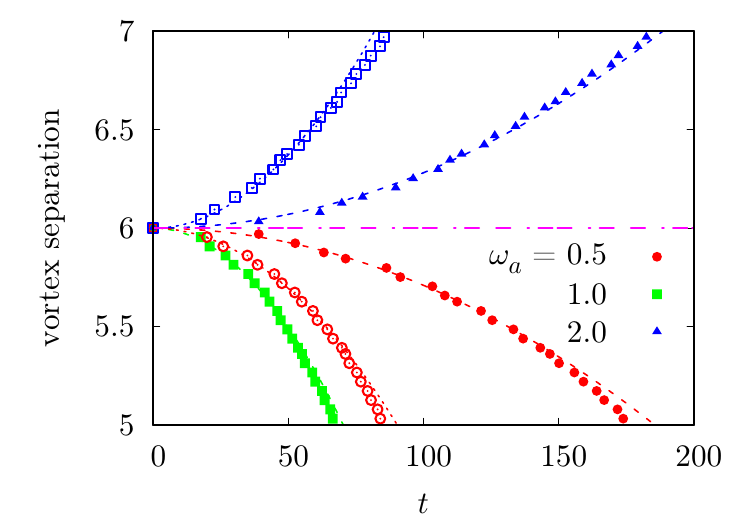}
\label{subfig:separation}
\includegraphics [width = 8.5cm, clip]{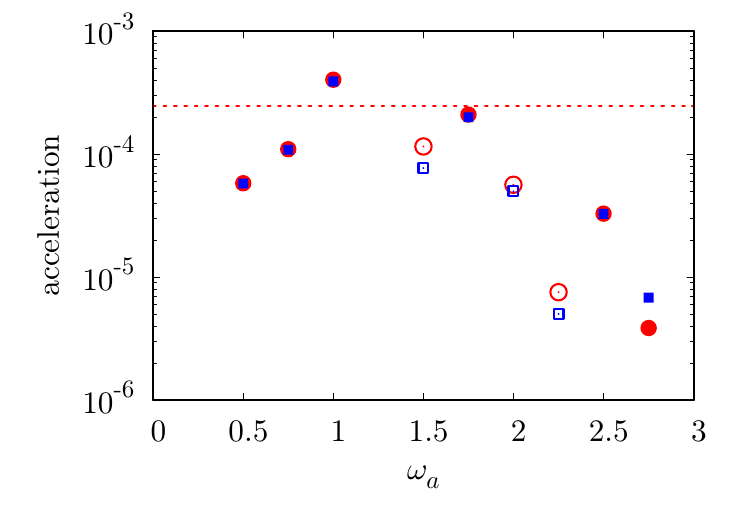}
\label{subfig:acceleration}
\caption{
Top: Time evolution of the separation between two vortices. 
Filled points represent the case with the critical coupling (BPS limit or $\beta = 1$) in the oscillating axion background with $\omega_a = 0.5$ (red circles), 1.0 (green squares), 2.0 (blue triangles) and open points represent the case without the axion but $\beta = 0.95$ (red circles), 1.1 (blue squares).
Data points are fitted by quadratic curves expressed by dashed or dotted lines. 
The horizontal dash-dotted line corresponds to the BPS case without axion-photon coupling. 
Bottom: Acceleration of a vortex. Filled (open) points represent positive (negative) values. The circle and square points respectively correspond to the case with $\mathcal{A} = 0.1$ and $0.15$, with the square points rescaled by $1.5^{-2}$, showing that the magnitude of the acceleration scales as $\mathcal{A}^2$.
As a reference, the horizontal dotted line shows the case with type-I vortex with $\beta = 0.95$ without the axion.
}
\label{fig:two_vortices}
\end{figure}

The upper panel of \FIG{fig:two_vortices} shows the time evolution of the separation $d$ of the two vortices
from the initial separation $d_{\rm ini}=6$.
The position of the vortex center is identified by the local winding number in a gauge-invariant way \cite{Kajantie:1998bg}.
For $\omega_a = 0.5$ and 1.0, the interaction between vortices is attractive, but for $\omega_a = 2.0$, vortices repel each other.
Those attractive/repulsive forces from the axion oscillation can be qualitatively understood by the interference between the induced oscillating electric and magnetic fields.
Since the induced electric field coherently oscillates, they are constructive with each other, so that the vortices repel to obtain the energetic stability.
On the other hand, the induced magnetic field oscillates circumferentially, and they are destructive as long as there is no overlap between the cores of vortices.  
When there is a significant overlap, the magnetic fields are no longer destructive, so that the interaction can be repulsive.

By sampling the data points satisfying $|d-d_{\rm ini}| < 0.5$,
we have calculated the acceleration associated with the attractive/repulsive forces between vortices, which is assumed to be constant in this interval.
In this case, the acceleration, $\alpha$, can be simply given by $d = d_{\rm ini} - \alpha t^2/2$ and the least squares fitting shows a good agreement with the data points as shown by dashed lines in the upper panel of \FIG{fig:two_vortices}.
Finally, the lower panel of \FIG{fig:two_vortices} shows the acceleration as a function of the axion frequency.
Filled and open points show the positive and negative values respectively, corresponding to the attractive and repulsive forces. 
The results show that the oscillating axion background can significantly change the dynamical feature of vortices. More detailed analysis is required to further investigate the quantitative nature of the vortex dynamics in the presence of the axion dark matter, which will be studied in the future work.

\vspace{2mm}
{\bf Discussion.--}
We have investigated the physics of the vortices in the Abelian-Higgs model with the oscillating axion background. 
In particular, the axion can induce oscillating electric fields inside the vortex. 
Moreover, such electric fields can be amplified if the oscillation frequency is tuned to some specific values. 
In this sense, the vortex can be regarded as a cylindrical resonant cavity. 
However, in the above analysis, we assume the vacuum inside the vortex core. 
It is not realistic in the superconductors, because there are a lot of free charged particles inside the vortex as in the ordinary conducting phase.
To realize such a realistic situation, we need to adopt a non-relativistic formulation including the electric conductivity in matter \cite{schmid1966time,kato1991computer}.

We have also shown that the background axion oscillation can qualitatively change the dynamics of two vortices. 
Namely, the axion induces both attractive and repulsive forces between vortices depending on its oscillation frequency. 
This feature can serve as a new basis for axion detection experiments.
So far, we have focused only on the BPS limit, but we are more interested in the type-II superconductor.
In this case, the formation of vortices can be enhanced by applying the external magnetic field, which lies between the higher and lower critical fields.
Thus, the extension of our analysis beyond the BPS limit is important for realistic experimental setups.

As a pioneering work, the idea of an axion probe based on stimulated axion emissions from superconducting vortices is proposed in \cite{Yokoi:2016umv}. Our findings can potentially propose a complementary and alternative way to probe the axion by focusing on the vortex dynamics itself. In particular, the axion can potentially change the physics of the phase transition caused by the creation of vortex pairs known as the Kosterlitz-Thouless transition \cite{Kosterlitz:1973xp}.
The possibility of axion detection with such phenomena will be investigated in the subsequent paper.

\vspace{2mm}
{\bf Acknowledgments.--}
We thank Satoshi Kashiwaya and Yasunori Mawatari for fruitful discussions.
This work used computational resources of Fugaku supercomputer, provided by RIKEN Center for Computational Sciences, through the HPCI System Research Project (Project ID: hp250177).

\bibliography{reference}

\end{document}